# ***IL-10*** **rs1800896 polymorphism predicts biochemical remission in IBD patients undergoing biologic therapy**

Michela Helga Falzone[1,†], Davide Giuseppe Ribaldone[1,2,†,*], Martina Buglione[1], Irene Cottone[1], Marta Vernero[2], Demis Pitoni[1], Angelo Armandi[1,2], Francesca Saba[1], Eleonora Dileo[1], Alfredo Santovito[3], Gian Paolo Caviglia[1]

[1]Department of Medical Sciences, University of Turin, Turin, Italy

[2]Division of Gastroenterology, Città della Salute e della Scienza – Molinette Hospital, Turin, Italy

[3]Department of Life Sciences and Systems Biology, University of Turin, Torino, Italy

[†]Co-first authors

*Corresponding author: Prof. Davide Giuseppe Ribaldone, Department of Medical Sciences, University of Turin, 10126, Turin, Italy. E-mail: davidegiuseppe.ribaldone@unito.it

**Conflict of interest**

The authors declare no potential conflicts of interest

**Funding**

This study was supported by the University of Turin under grant numbers RIBD_RILO_22_01 and CAVG_RILO_23_01.

**Data sharing statement**

Data are available upon reasonable request to the corresponding author.

Word count: 2428

Number of Tables: 2

Number of Figures: 1

Number of Supplementary Tables: 6

Number of Supplementary Figures: 1

**Abstract**

**Background.** Genetic factors, including single-nucleotide polymorphisms (SNPs), may modulate disease course and therapeutic efficacy in patients with inflammatory bowel disease (IBD).

**Aim.** We investigated the association between four SNPs in cytokine genes and clinical phenotype as well as the response to molecular-targeted drugs in patients with IBD.

**Materials and Methods.** A total of 197 IBD patients (142 Crohn’s disease [CD], 55 ulcerative colitis [UC]) undergoing targeted treatment were enrolled. The SNPs analyzed were: *TNF-α* rs1800629 (-308 G>A), *TGF-β* rs1800471 (-codon 10 C>T), *IL-6* rs1800795 (-174 G>C), and *IL-10* rs1800896 (-1082 G>A). Biochemical response at 12 months (T12) was defined as CRP <5.0 mg/L and FC <250 µg/g at T12, in the absence of ongoing corticosteroid therapy.

**Results.** Among the SNPs analyzed, the *IL-6* rs1800795 C allele was significantly associated with a younger age at diagnosis (p = 0.049), while the *TNF-α* rs1800629 A allele was more frequently observed in patients with CD than in UC (p = 0.036). Regarding treatment response, 134 patients completed 12 months of molecular-targeted therapy and were included in the per-protocol analysis; 41.0% achieved biochemical remission at T12. The *IL-10* rs1800896 mutated allele was significantly associated with remission (OR = 2.15, 95% CI 1.03–4.44; p = 0.041). This association remained significant in multivariate analysis (aOR = 4.15, 95% CI 1.49–11.56; p = 0.007), independently of clinical and treatment-related variables.

**Conclusions.** Genotyping of cytokine-related SNPs may help identify patients with a more aggressive disease phenotype and guide personalized treatment strategies in patients with IBD.

**Key words:** Crohn’s disease; cytokine; inflammatory bowel disease; single-nucleotide polymorphisms; ulcerative colitis; infliximab, adalimumab, vedolizumab, ustekinumab, tofacitinib.

## Introduction

Inflammatory Bowel Disease (IBD), encompassing ulcerative colitis (UC) and Crohn's disease (CD), affects approximately 7 million individuals worldwide. Over the past 30 years, the incidence and prevalence of IBD have risen significantly, particularly among individuals aged 15-45 years. While IBD remains more prevalent in Western countries, the incidence is rising in low- and middle-income countries as well, likely due to the interaction of genetic predisposition, environmental factors, and lifestyle changes related to urbanization (1).

IBD requires lifelong management, traditionally following a step-up approach from conventional treatments to advanced therapies. This results in a substantial economic burden due to the cumulative costs of long-term pharmacotherapy, outpatient care, diagnostics, and hospitalizations during flares (2). Despite the availability of multiple biologic agents over the past two decades (3), current guidelines lack a defined therapeutic sequence, underscoring the need for strategies that optimize efficacy, safety, and cost-effectiveness (4).

Given these challenges, increasing attention has been directed toward identifying biomarkers to inform therapeutic decisions and personalize care. Several genetic variants have been investigated in relation to IBD phenotypes and, more recently, to therapeutic outcomes; a substantial proportion of these variants are located within genes encoding cytokines (5), such as interleukin (IL)-6, IL-10, tumor necrosis factor (TNF)-α, and transforming growth factor (TGF)-β, which play a central role in IBD development and progression. Consistent with this, recent studies have shown that circulating levels of these cytokines may be useful to predict response to biologics, including anti-TNF drugs, vedolizumab (VDZ), and ustekinumab (UST) (6,7).

In the present study, we aimed to investigate whether specific polymorphisms in cytokine-related genes, namely *IL-6* rs1800795 (-174 G>C), *IL-10* rs1800896 (-1082 G>A), *TNF-α* rs1800629 (-308 G>A), and *TGF-β* rs1800471 (codon 10 T>C), are associated with disease phenotype and biochemical response to targeted therapies in patients with IBD. These cytokine-gene variants were selected based on prior evidence of functional effects on gene transcription or protein expression, and their relevance to pathways central to IBD pathogenesis and therapeutic response. *IL-6* rs1800795 has been linked to inflammatory activity and disease susceptibility (8), while *IL-10* rs1800896 is known to modulate IL-10 expression and has been associated with IBD risk in population and meta-analytic studies (9). *TNF-α* rs1800629 influences TNF expression and has been widely investigated, albeit with inconsistent results, in relation to anti-TNF response (10). Finally, the *TGF-β* rs1800471 affects cytokine production and is increasingly recognized in pharmacogenetic models of IBD (11). Together, these variants represent a biologically plausible panel to explore the contribution of host cytokine genetics to disease phenotype and therapeutic outcomes in IBD.

## Materials and Methods

### *Study design*

We conducted an ambispective, observational, single-center study at the Division of Gastroenterology of “AOU Città della Salute e della Scienza di Torino”, Italy, enrolling consecutive patients with IBD undergoing treatment with targeted therapies between January 2018 and June 2024.

Eligible participants were patients aged ≥ 18 years with a confirmed diagnosis of CD or UC according to the ECCO-ESGAR guidelines (12). All patients were initiating molecular-targeted therapies, including infliximab (IFX), adalimumab (ADA), VDZ, UST, and tofacitinib (TOF), and provided written informed consent to participate in the study and undergo genetic testing. The cytokine gene polymorphisms analyzed are listed in Supplementary Table 1.

Demographic, clinical, and biochemical data were retrospectively collected from medical records. These included: age at treatment initiation, sex, smoking status (including former smokers), disease type (CD or UC), disease duration, prior surgical interventions, and disease localization according to the Montreal classification (13). CD localization was classified as ileal (L1), colonic (L2), ileocolonic (L3), or, as additional parameter, upper gastrointestinal involvement (L4); UC extent was defined as proctitis (E1), left-sided colitis (E2), or extensive colitis (E3). Clinical disease activity was assessed using the Harvey–Bradshaw Index for CD and the Partial Mayo Score for UC (14,15). Biochemical parameters, including C-reactive protein (CRP) and fecal calprotectin (FC), were collected at baseline (T0) and 12 months of treatment (T12).

The study was conducted in accordance with the Declaration of Helsinki and was approved by the local Ethics Committee: Comitato Etico Interaziendale A.O.U. Città della Salute e della Scienza di Torino — A.O. Ordine Mauriziano — A.S.L. Città di Torino (approval code n. 0056924; date: 8-jun-2016).

### *Study endpoints*

The primary endpoint of the study was to assess the association between selected cytokine gene polymorphisms and clinical characteristics of IBD, including disease type (CD vs. UC), age at diagnosis, and history of surgical interventions.

Secondary endpoints included the evaluation of correlations between these polymorphisms and the biochemical response at T12 to molecular-targeted therapies, with a specific focus on patients with CD and anti-TNF agents. Biochemical response was defined as CRP < 5 mg/L and FC < 250 μg/g at T12, in the absence of ongoing corticosteroid therapy (16).

*Genetic analysis*

Whole blood samples were collected in EDTA tubes, centrifuged at 8000 rpm, and stored in polypropylene tubes at −80 °C until analysis. Genomic DNA was isolated using an automated protocol based on magnetic particle technology (QIAGEN GmbH, Hilden, Germany) with the Qiagen EZ1 Advanced XL instrument, following the manufacturer's instructions.

For the analysis of *IL-6* rs1800795, *IL-10* rs1800896, *TNF-α* rs1800629, and *TGF-β* rs1800471, polymorphisms, allelic discrimination was assessed through Amplification Refractory Mutation System (ARMS)-PCR methodology, with primers listed in Supplementary Table 1. PCR reactions were carried out in a final volume of 25 μL, containing approximately 10 ng of genomic DNA, 1x Reaction Buffer, 1.5 mM $MgCl_2$, 5% DMSO, 250 μM dNTPs, 0.5 μM of each primer, and 1 U of Taq DNA polymerase (all reagents from Thermo Fisher Scientific, Waltham, MA, USA). The thermal cycling conditions included 35 cycles of denaturation at 95 °C for 1 min, annealing at 60 °C for 1 min, and extension at 72 °C for 1 min, followed by a final extension at 72 °C for 10 min. PCR products were visualized by electrophoresis on 3% agarose gels (Società Italiana Chimici, Rome, Italy), stained with ethidium bromide (Merck, Milan, Italy).

*Statistical analysis*

Categorical variables were expressed as absolute frequencies and percentages. The normality of continuous variables was assessed using the Shapiro–Wilk test. As none of the variables followed a normal distribution, results were presented as median and interquartile range (IQR), and comparisons between groups were performed using the Mann–Whitney U test.

Differences in the distribution of cytokine SNPs between groups were assessed using the chi-square ($\chi^2$) test. The association between SNPs and biochemical remission at T12 was analyzed through logistic regression, with results expressed as odds ratios (OR) and 95% confidence intervals (CI). A p-value < 0.05 was considered statistically significant. All the statistical analyses were performed with MedCalc software, version 20.1 (MedCalc, Ostend, Belgium).

## Results

*Study population*

A total of 197 patients with IBD were enrolled in the study, including 142 (72.1%) with CD and 55 (27.9%) with UC. Most patients were males (n = 133; 67.5%); median age at treatment initiation was 45 (33–59) years. The median age at IBD diagnosis was 28 years (21–40), and 81 patients had already undergone at least one surgical intervention. At baseline, 50 (25.4%) patients were in clinical remission, 66 (33.6%) had mild disease, and 81 (41.1%) presented with moderate to severe disease activity; 112 patients were under systemic corticosteroid treatment. Additional patient characteristics are reported in Table 1.

*Association between SNPs and patients' baseline features*

Genotype distributions for each polymorphism in the study cohort are presented in Supplementary Table 2. No statistically significant differences were observed in the genotypic frequencies of *IL-6* rs1800795 and *IL-10* rs1800896 genotypes between CD and UC patients. However, a significant difference was found for *TNF-α* rs1800629 ($p = 0.020$): the wild-type (wt) GG genotype was more prevalent in UC patients (78.2%) compared to those with CD (62.4%), while the presence of the A allele (either GA or AA genotypes) was more frequent in patients with CD ($p = 0.036$) (Supplementary Table 3).

Among the polymorphisms analyzed, only *IL-6* rs1800795 showed a statistically significant association with age at diagnosis. Patients with the GG genotype had a median age at diagnosis of 30 (22–41) years, whereas carriers of at least one mutated allele (GC or CC) had an earlier onset, with a median age of 25 (19–36) years ($p = 0.049$).

Finally, no statistically significant associations were found between any SNP and disease extension (Supplementary Table 4) and history of prior surgery (Supplementary Table 5). However, patients carrying at least one mutated allele of *IL-6* rs1800795 were more likely to have undergone surgery compared to those with the wt genotype, although this difference did not reach statistical significance ($p = 0.076$).

*Association between SNPs and treatment outcome*

Patients in the study underwent molecular-targeted therapy, with the majority treated with anti-TNF agents, specifically IFX (n = 16; 8.1%) and ADA (n = 90; 45.7%). Other therapies included VDZ, n = 50; 25.4%), UST (n = 35; 17.8%), and TOF (n = 6; 3.0%). Out of the initial 197 patients, 134 completed 12 months of follow-up, and had available biochemical data and were therefore included in the per-protocol analysis. Among these, biochemical remission at T12 was achieved in 55 (41.0%) of patients.

Genotype distributions according to treatment outcome are reported in Supplementary Table 6. Under a dominant genetic model, a significantly different distribution of *IL-10* rs1800896 genotypes was observed between patients who achieved biochemical remission at T12 and those who did not (Figure 1); the presence of at least one mutated allele (GA or AA) was significantly associated with a higher likelihood of biochemical remission (OR = 2.15, 95% CI 1.03–4.44; p = 0.041). This association remained significant in a sensitivity analysis limited to patients treated with anti-TNF agents (n = 106; OR = 3.36, 95% CI 1.07–10.58; p = 0.034) as well as in patients with CD regardless of treatment regimen (n = 98; OR = 3.06, 95% CI 1.28–7.27; p = 0.011) (Supplementary Figure 1).

Finally, multivariate logistic regression analysis confirmed that the presence of at least one mutated allele in the *IL-10* rs1800896 SNP was independently associated with biochemical remission at T12 (aOR = 4.15, 95% CI 1.49–11.56; p = 0.007) after adjustment for age, sex, disease type, duration and activity, prior surgery, and type of molecular-targeted therapy (Table 2).

**Discussion**

This study explored the association between specific cytokine-related SNPs and both disease characteristics and therapeutic response in patients with IBD. Among the four polymorphisms analyzed, *IL-6* rs1800795 (-174 G>C), *IL-10* rs1800896 (-1082 G>A), *TNF-α* rs1800629 (-308 G>A), and *TGF-β* rs1800471 (codon 10 T>C), we observed that the C allele of *IL-6* rs1800795 C was significantly associated with younger age at diagnosis, while the A allele of *TNF-α* rs1800629 was more frequently detected among patients with CD compared to those with UC. Notably, *IL-10* rs1800896 emerged as the most clinically relevant variant, demonstrating a strong and independent association between the mutated allele and biochemical remission at 12 months in patients treated with molecular-targeted therapies. These results support the potential utility of cytokine gene polymorphisms as predictive biomarkers in the context of personalized medicine for IBD.

Among the SNPs analyzed according to baseline clinical features, *IL-6* rs1800795 and *TNF-α* rs1800629 displayed modest but potentially meaningful associations with disease characteristics. The presence of mutated C allele of *IL-6* rs1800795 was significantly associated with a younger age at diagnosis, and a trend toward increased history of surgery was observed, suggesting a possible link between this SNP and a more aggressive disease phenotype. This finding is consistent with prior reports of elevated IL-6 and CRP levels in CC genotype carriers and increased clinical severity in UC patients (8). Regarding *TNF-α* rs1800629, we found that the A allele was more frequent in CD than UC, although no association with treatment response was observed. Available data on this polymorphism are conflicting, with some studies reporting significant associations and others showing no effect (10,17,18). The A allele at *TNF-α* rs1800629 occurs within the *TNF-α* promoter and has been associated with higher *TNF-α* transcriptional activity and with IBD susceptibility/phenotype differences. In our cohort, its higher frequency in CD than UC is therefore biologically plausible, given the stronger TNF-driven, Th1-skewed inflammation typically seen in CD. At the same time, prior studies evaluating *TNF-α* rs1800629 in IBD reported heterogeneous effects on disease risk and treatment outcomes (19-21), with several cohorts finding no robust predictive value for anti-TNF response (mirrored by our data). Taken together, these results suggest that *TNF-α* rs1800629 may contribute modestly to disease phenotype (CD vs. UC), but alone is unlikely to serve as a reliable therapeutic biomarker without larger, stratified analyses and standardized endpoints. These discrepancies may be due to differences in study populations, definitions of therapeutic response, and the specific biologic agents investigated. Taken together, these results suggest that both variants may contribute to delineating patient subgroups with distinct disease phenotypes, although their clinical relevance needs to be further confirmed.

*IL-10* rs1800896 was the only SNP in our analysis to show a clear association with treatment response. Carriers of the A allele were more likely to achieve biochemical remission at 12 months following molecular-targeted therapy, and this association remained consistent across the subgroup analyses and multivariate model. IL-10 is involved in the regulation of intestinal inflammation (22), and variations in its promoter region have been shown to influence cytokine expression levels (9). The A allele at position -1082 has been associated with altered transcriptional activity and may affect the anti-inflammatory balance in IBD. Notably, a 2020 meta-analysis by Su et al. linked this SNP to increased susceptibility to IBD development (23), highlighting its potential biological relevance. However, to the best of our knowledge, our study is the first to report a significant association between *IL-10* rs1800896 and response to molecular-targeted therapies in patients with IBD. While these findings are promising, they should be interpreted cautiously and warrant validation in larger, independent populations before being translated into clinical practice.

Some study limitations should be acknowledged. First, due to the ambispective nature of the study, not all patients were included in the follow-up analyses. Second, treatment response was assessed using biochemical markers (CRP and FC), which, although widely used and clinically relevant, do not fully reflect mucosal healing; endoscopic assessment would have provided a more solid measure of therapeutic efficacy. Third, the relatively small number of patients in some treatment subgroups, particularly those receiving VDZ, UST, or TOF, may have limited the power of analyses.

In conclusion, our findings suggest that *IL-6* rs1800795 and *TNF*-α rs1800629 may contribute to defining distinct clinical phenotypes in IBD, potentially reflecting more aggressive disease courses. In parallel, *IL-10* rs1800896 emerged as a promising genetic marker of a favorable response to molecular-targeted therapies. Together, these results provide novel insights into the role of host genetic variability in shaping both disease expression and treatment outcomes, supporting the development of precision medicine approaches for the management of patients with IBD.

**Table 1.** Patients' characteristics at treatment initiation (T0).

| **Variables** | |
|---|---|
| Male sex, n (%) [197] | 133 (67.5%) |
| Age, median (IQR) [197] | 45 (33–59) |
| Age at diagnosis, median (IQR) [197] | 28 (21–40) |
| Family history, n (%) [190] | 32 (16.8%) |
| Active smokers, n (%) [196] | 53 (27.0%) |
| Ex-smokers, n (%) [196] | 64 (32.7%) |
| Past surgical intervention, n (%) [196] | 81 (41.3%) |
| CD vs. UC, n (%) [197] | 142 (72.1%)/55 (27.9%) |
| Extent of CD [142] | |
| - L1, n (%) | 43 (30.3%) |
| - L2, n (%) | 8 (5.6%) |
| - L3, n (%) | 86 (60.6%) |
| - L1 + L4, n (%) | 1 (0.7%) |
| - L2 + L4, n (%) | 2 (1.4%) |
| - L3 + L4, n (%) | 2 (1.4%) |
| Extent of UC [54] | |
| - E1, n (%) | 6 (11.1%) |
| - E2, n (%) | 19 (35.2%) |
| - E3, n (%) | 29 (53.7%) |
| Clinical disease activity [197] | |
| - Remission, n (%) | 50 (25.4%) |
| - Mild, n (%) | 66 (33.6%) |
| - Moderate, n (%) | 67 (34.0%) |
| - Severe, n (%) | 14 (7.1%) |
| Systemic corticosteroids, n (%) [196] | 112 (57.1%) |
| CRP (mg/L), median (IQR) [177] | 6.8 (2.6–18.3) |
| FC (μg/g), median (IQR) [146] | 518 (158–1500) |

Numbers in brackets indicate the number of patients with available data.

Abbreviations: CD, Crohn’s disease; CRP, C-reactive protein; FC, fecal calprotectin; IQR, interquartile range; n, number; T0, baseline; UC, ulcerative colitis.

**Table 2.** Multivariate logistic regression analysis of predictors of biochemical remission at T12.

| **Variables** | **aOR (95% CI)** | **p-value** |
|---|---|---|
| Age (years) | 1.02 (0.97–1.07) | 0.550 |
| Male sex | 1.37 (0.52–3.61) | 0.530 |
| UC diagnosis | 0.44 (0.12–1.71) | 0.237 |
| Disease duration (years) | 0.99 (0.94–1.03) | 0.604 |
| Previous surgery | 1.02 (0.37–2.79) | 0.976 |
| Disease activity (T0) | 0.96 (0.85–1.09) | 0.540 |
| anti-TNFα treatment | 2.84 (0.88–9.19) | 0.081 |
| *IL-10* rs1800896 GA + AA | 4.15 (1.49–11.56) | 0.007 |

Abbreviations: aOR, adjusted odds ratio; CI, confidence interval; IL-10, interleukin-10; UC, ulcerative colitis.

**Figure 1.** Distribution of *IL-10* rs1800896 genotypes according to biochemical remission at T12.

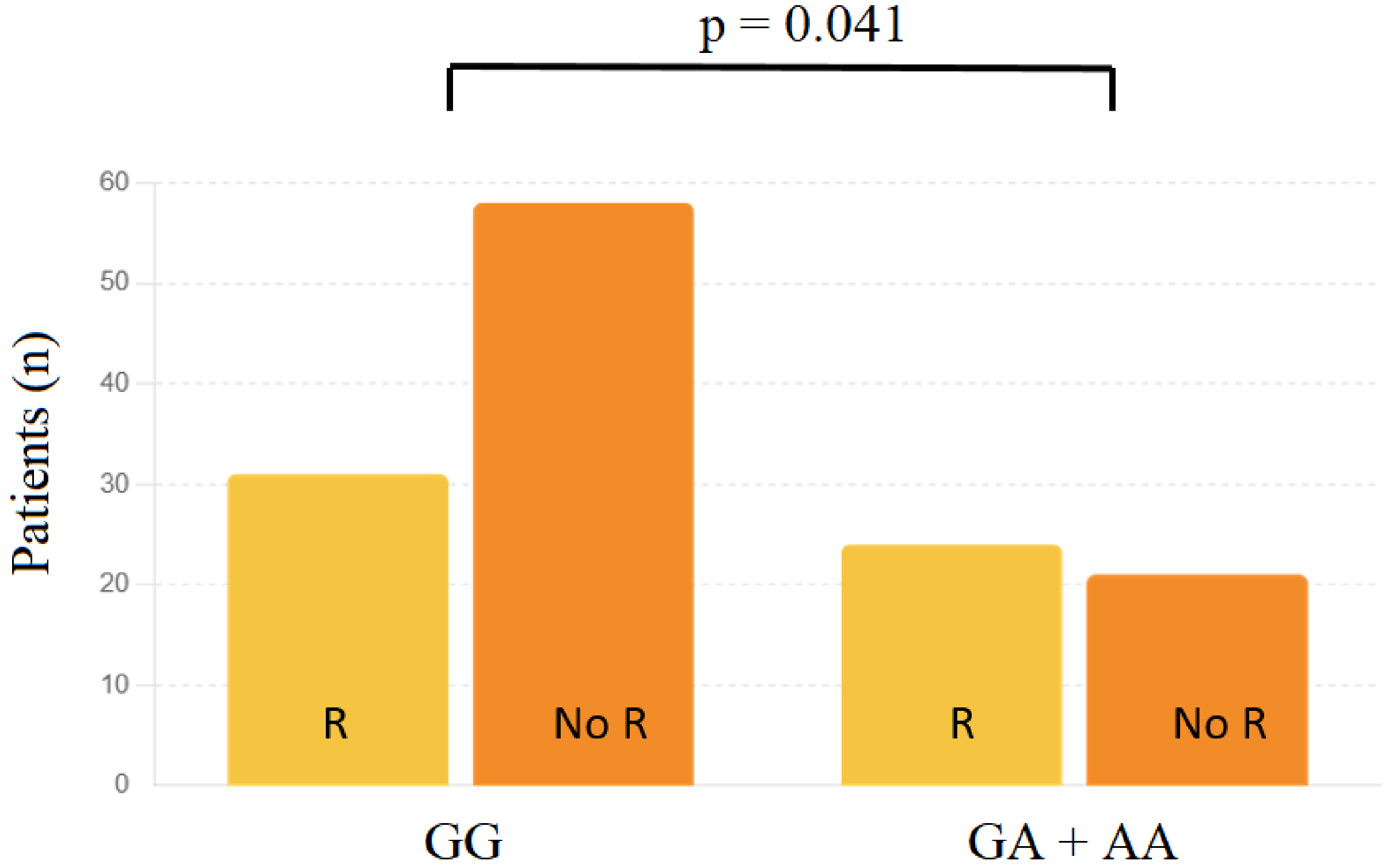

Dominant genetic model: individuals with at least one copy of the variant allele (heterozygous or homozygous) compared to wild-type homozygotes.

Abbreviations: n, number; NR, non-remission; R, remission; T12, month 12.

**Supplementary Table 1.** Primers and PCR products for gene polymorphisms analyzed in the present study. For all genes, the annealing temperature was 60°C and the used methodology was Amplification Refractory Mutation System PCR (ARMS-PCR).

| Gene | Reference SNP | Sequence | PCR product (bp) |
|---|---|---|---|
| *IL-6* (-174, G>C) | rs1800795 | | 190 |
| - Antisense primer | | 5'-TCGTGCATGACTTCAGCTTTA-3' | |
| - G-sense primer | | 5'-AATGTGACGTCCTTTAGCATG-3' | |
| - C-sense primer | | 5'-AATGTGACGTCCTTTAGCATC-3' | |
| *IL 10* (-1082, G>A) | rs1800896 | | 258 |
| - Antisense primer | | 5'-AGTGCCAACTGAGAATTTGG-3' | |
| - G-sense primer | | 5'-CTACTAAGGCTTCTTTGGGAG-3' | |
| - A-sense primer | | 5'-ACTACTAAGGCTTCTTTGGGAA-3' | |
| *TNF-α* (-308, G>A) | rs1800629 | | 184 |
| - Antisense primer | | | |
| - G-sense primer | | 5'-ATAGGTTTTGAGGGGCATGG-3' | |
| - A-sense primer | | 5'-AATAGGTTTTGAGGGGCATGA-3' | |
| *TGF-β* (Codon 10, T>C) | rs1800471 | | 241 |
| - Antisense primer | | 5'-TCCGTGGGATACTGAGACAC-3' | |
| - C-sense primer | | 5'-GCAGCGGTAGCAGCAGCG-3' | |
| - T-sense primer | | 5'-AGCAGCGGTAGCAGCAGCA-3' | |

Abbreviations: bp, base pair; IL-6, interleukin-6; IL-10, interleukin-10; SNPs, single-nucleotide polymorphisms; TGF-β, transforming growth factor (TGF)-β; TNF-α, tumor necrosis factor-α.

**Supplementary Table 2.** Distribution of *IL-6* rs1800795, *IL-10* rs1800896, *TNF-α* rs1800629, and *TGF-β* rs1800471 in the study cohort.

| SNPs | n (%) |
|---|---|
| *IL-6* rs1800795 (-174 G>C) [197] | |
| - GG, n (%) | 134 (68.0%) |
| - GC, n (%) | 58 (29.4%) |
| - CC, n (%) | 5 (2.5%) |
| *IL-10* rs1800896 (-1082 G>A) [197] | |
| - GG, n (%) | 135 (68.5%) |
| - GA, n (%) | 55 (27.9%) |
| - AA, n (%) | 7 (3.6%) |
| *TNF-α* rs1800629 (-308 G>A) [196] | |
| - GG, n (%) | 131 (66.8%) |
| - GA, n (%) | 59 (30.1%) |
| - AA, n (%) | 6 (3.1%) |
| *TGF-β* rs1800471 (codon 10 T>C) [197] | |
| - TT, n (%) | 136 (69.0%) |
| - TC, n (%) | 59 (29.9%) |
| - CC, n (%) | 2 (1.0%) |

*TNF-α* rs1800629 (-308 G>A) polymorphism was analyzed in 196 patients, while genotyping for *IL-6* rs1800795 (-174 G>C), *IL-10* rs1800896 (-1082 G>A), and *TGF-β* rs1800471 (codon 10 C>T) was available for 197 patients.

Abbreviations: IL-6, interleukin-6; IL-10, interleukin-10; n, number; SNPs, single-nucleotide polymorphisms; TGF-β, transforming growth factor (TGF)-β; TNF-α, tumor necrosis factor-α.

**Supplementary Table 3.** Distribution of *IL-6* rs1800795, *IL-10* rs1800896, *TNF-α* rs1800629, and *TGF-β* rs1800471 in the study cohort according to diagnosis of CD or UC.

| SNPs | CD | UC | p-value |
|---|---|---|---|
| *IL-6* rs1800795 (-174 G>C) [197] | | | |
| - GG, n (%) | 95 (66.9%) | 39 (70.9%) | 0.549 |
| - GC, n (%) | 43 (30.3%) | 15 (27.3%) | |
| - CC, n (%) | 4 (2.8%) | 1 (1.8%) | |
| *IL-10* rs1800896 (-1082 G>A) [197] | | | |
| - GG, n (%) | 96 (67.6%) | 39 (70.9%) | 0.831 |
| - GA, n (%) | 43 (30.3%) | 12 (21.8%) | |
| - AA, n (%) | 3 (2.1%) | 4 (7.3%) | |
| *TNF-α* rs1800629 (-308 G>A) [196] | | | |
| - GG, n (%) | 88 (62.4%) | 43 (78.2%) | 0.020 |
| - GA, n (%) | 47 (33.3%) | 12 (21.8%) | |
| - AA, n (%) | 6 (4.3%) | 0 | |
| *TGF-β* rs1800471 (codon 10 T>C) [197] | | | |
| - TT, n (%) | 99 (69.7%) | 37 (67.3%) | 0.649 |
| - TC, n (%) | 42 (29.6%) | 17 (30.9%) | |
| - CC, n (%) | 1 (0.7%) | 1 (1.8%) | |

Numbers in brackets indicate the number of patients with available data.

Abbreviations: CD, Crohn's disease; IL-6, interleukin-6; IL-10, interleukin-10; n, number; SNPs, single-nucleotide polymorphisms; TGF-β, transforming growth factor (TGF)-β; TNF-α, tumor necrosis factor-α; UC, ulcerative colitis.

**Supplementary Table 4.** Distribution of *IL-6* rs1800795, *IL-10* rs1800896, *TNF-α* rs1800629, and *TGF-β* rs1800471 in the study cohort according to disease extension.

| | CD extension | | | UC extension | | |
|---|---|---|---|---|---|---|
| **SNPs** | **L2** | **Other** | **p-value** | **E3** | **Other** | **p-value** |
| *IL-6* rs1800795 (-174 G>C) [197] | | | | | | |
| - GG, n (%) | 7 (70.0%) | 88 (66.7%) | 0.717 | 22 (75.9%) | 17 (68.0%) | 0.380 |
| - GC, n (%) | 3 (30.0%) | 40 (30.3%) | | 7 (24.1%) | 7 (28.0%) | |
| - CC, n (%) | 0 | 4 (3.0%) | | 0 | 1 (4.0%) | |
| *IL-10* rs1800896 (-1082 G>A) [197] | | | | | | |
| - GG, n (%) | 7 (70.0%) | 89 (67.4%) | 0.775 | 21 (72.4%) | 17 (68.0%) | 0.743 |
| - GA, n (%) | 3 (30.0%) | 40 (30.3%) | | 6 (20.7%) | 6 (24.0%) | |
| - AA, n (%) | 0 | 3 (2.3%) | | 2 (6.9%) | 2 (8.0%) | |
| *TNF-α* rs1800629 (-308 G>A) [196] | | | | | | |
| - GG, n (%) | 6 (60.0%) | 82 (62.6%) | 0.641 | 23 (79.3%) | 19 (76.0%) | 0.773 |
| - GA, n (%) | 3 (30.0%) | 44 (33.6%) | | 6 (20.7%) | 6 (24.0%) | |
| - AA, n (%) | 1 (10.0%) | 5 (3.8%) | | 0 | 0 | |
| *TGF-β* rs1800471 (codon 10 T>C) [197] | | | | | | |
| - TT, n (%) | 5 (50.0%) | 94 (71.2%) | 0.192 | 20 (69.0%) | 16 (64.0%) | 0.523 |
| - TC, n (%) | 5 (50.0%) | 37 (28.0%) | | 9 (31.0%) | 8 (32.0%) | |
| - CC, n (%) | 0 | 1 (0.8%) | | 0 | 1 (4.0%) | |

Numbers in brackets indicate the number of patients with available data.

Abbreviations: CD, Crohn's disease; IL-6, interleukin-6; IL-10, interleukin-10; n, number; SNPs, single-nucleotide polymorphisms; TGF-β, transforming growth factor (TGF)-β; TNF-α, tumor necrosis factor-α; UC, ulcerative colitis.

**Supplementary Table 5.** Distribution of *IL-6* rs1800795, *IL-10* rs1800896, *TNF-α* rs1800629, and *TGF-β* rs1800471 in the study cohort according to previous surgery.

| | Previous surgery | | |
|---|---|---|---|
| **SNPs** | **Yes** | **No** | **p-value** |
| *IL-6* rs1800795 (-174 G>C) [197] | | | |
| - GG, n (%) | 52 (64.2%) | 82 (71.3%) | 0.143 |
| - GC, n (%) | 25 (30.9%) | 32 (27.8%) | |
| - CC, n (%) | 4 (4.9%) | 1 (0.9%) | |
| *IL-10* rs1800896 (-1082 G>A) [197] | | | |
| - GG, n (%) | 52 (64.2%) | 82 (71.3%) | 0.146 |
| - GA, n (%) | 24 (29.6%) | 31 (27.0%) | |
| - AA, n (%) | 5 (6.2%) | 2 (1.7%) | |
| *TNF-α* rs1800629 (-308 G>A) [196] | | | |
| - GG, n (%) | 54 (66.7%) | 76 (66.7%) | 0.686 |
| - GA, n (%) | 23 (28.4%) | 36 (31.6%) | |
| - AA, n (%) | 4 (4.9%) | 2 (1.8%) | |
| *TGF-β* rs1800471 (codon 10 T>C) [197] | | | |
| - TT, n (%) | 61 (75.3%) | 74 (64.3%) | 0.135 |
| - TC, n (%) | 19 (23.5%) | 40 (34.8%) | |
| - CC, n (%) | 1 (1.2%) | 1 (0.9%) | |

Numbers in brackets indicate the number of patients with available data.

Abbreviations: IL-6, interleukin-6; IL-10, interleukin-10; n, number; SNPs, single-nucleotide polymorphisms; TGF-β, transforming growth factor (TGF)-β; TNF-α, tumor necrosis factor-α.

**Supplementary Table 6.** Distribution of *IL-6* rs1800795, *IL-10* rs1800896, *TNF-α* rs1800629, and *TGF-β* rs1800471 in the study cohort according to biochemical remission at T12.

| | **Biochemical remission at T12** | | |
|---|---|---|---|
| **SNPs** | **Yes** | **No** | **p-value** |
| *IL-6* rs1800795 (-174 G>C) [134] | | | |
| - GG, n (%) | 38 (69.1%) | 50 (63.3%) | 0.542 |
| - GC, n (%) | 15 (27.3%) | 26 (32.9%) | |
| - CC, n (%) | 2 (3.6%) | 3 (3.8%) | |
| *IL-10* rs1800896 (-1082 G>A) [134] | | | |
| - GG, n (%) | 31 (27.0%) | 58 (73.4%) | 0.211 |
| - GA, n (%) | 23 (41.8%) | 16 (20.3%) | |
| - AA, n (%) | 1 (1.8%) | 5 (6.3%) | |
| *TNF-α* rs1800629 (-308 G>A) [133] | | | |
| - GG, n (%) | 36 (66.7%) | 54 (68.4%) | 0.545 |
| - GA, n (%) | 16 (29.6%) | 25 (31.6%) | |
| - AA, n (%) | 2 (3.7%) | 0 | |
| *TGF-β* rs1800471 (codon 10 T>C) [134] | | | |
| - TT, n (%) | 40 (72.7%) | 52 (65.8%) | 0.470 |
| - TC, n (%) | 14 (25.5%) | 26 (32.9%) | |
| - CC, n (%) | 1 (1.8%) | 1 (1.3%) | |

Numbers in brackets indicate the number of patients with available data.

Abbreviations: IL-6, interleukin-6; IL-10, interleukin-10; n, number; SNPs, single-nucleotide polymorphisms; TGF-β, transforming growth factor (TGF)-β; TNF-α, tumor necrosis factor-α.

**Supplementary Figure 1.** Distribution of *IL-10* rs1800896 genotypes according to biochemical remission at T12 in (A) patients treated exclusively with anti-TNF agents, and (B) patients with CD regardless of treatment regimen.

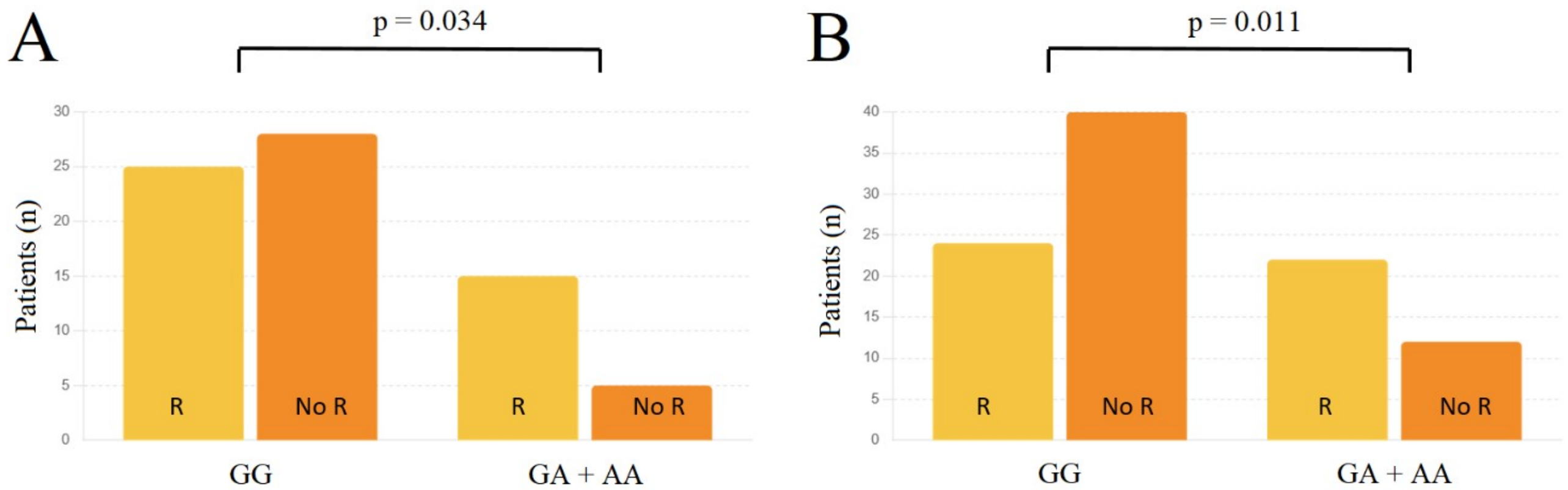

Dominant genetic model: individuals with at least one copy of the variant allele (heterozygous or homozygous) compared to wild-type homozygotes.

Abbreviations: CD, Crohn's disease; n, number; NR, non-remission; R, remission; T12, month 12.